\documentclass[preprint,aps]{revtex4}
\usepackage{graphicx}
\def\Vec#1{\mbox{\boldmath $#1$}}

\begin{document}

\title{Internal One-Particle Density Matrix for Bose-Einstein Condensates with Finite Number of Particles in a Harmonic Potential}

\author{Taiichi Yamada, Yasuro Funaki$^1$, Hisashi Horiuchi$^{2,3}$, Gerd R\"opke$^4$, Peter Schuck$^{5,6,7}$, and Akihiro\ Tohsaki$^2$ }

\affiliation{Laboratory of Physics, Kanto Gakuin University, Yokohama 236-8501, Japan}

\affiliation{$^{1}$Nishina Center of Accelerator-based Sciences, The Institute of Physical and Chemical Research (RIKEN), Wako 351-0098, Japan}

\affiliation{$^{2}$ Research Center for Nuclear Physics, Osaka\ University, Osaka 567-0047, Japan}

\affiliation{$^{3}$ International Institute for Advanced Studies, Kizugawa 619-0225,Japan}

\affiliation{$^{4}$ Institut f\"ur Physik, Unversit\"at Rostock, D-18051 Rostock, Germany}

\affiliation{$^{5}$ Institut de Physique Nucl\'eaire, CNRS, UMR 8608, Orsay, F-91505, France}

\affiliation{$^{6}$ Universit\'e Paris-Sud, Orsay, F-91505, France}

\affiliation{$^{7}$ Laboratoire de Physique et Mod\'elisation des Milieux Condens\'es, CNR et Universit\'e Joseph Fourier,  
             25 Av. des Martyrs, BP 166, F-38042 Grenoble Cedex 9, France }

\begin{abstract}

Investigations on the internal one-particle density matrix in the case 
 of Bose-Einstein condensates with a finite number ( $N$ ) of particles 
 in a harmonic potential are performed. 
We solve the eigenvalue problem of the Pethick-Pitaevskii-type internal
 density matrix and find a fragmented condensate.  
On the contrary the condensate Jacobi-type internal density matrix 
 gives complete condensation into a single state.
The internal one-particle density matrix is, therefore, shown to be different
 in general for different choices of the internal coordinate system. 
We propose two physically motivated criteria for the choice of 
 the adequate coordinate systems which give us a unique answer for the 
 internal one-particle density matrix. 
One criterion is that in the infinite particle number limit ( $N=\infty$ )
 the internal one-particle density matrix should have the same eigenvalues
 and eigenfunctions as those of the corresponding ideal Bose-Einstein condensate
 in the laboratory frame. 
The other criterion is that the coordinate of the internal one-particle
 density matrix should be orthogonal to the remaining $( N - 2 )$ internal
 coordinates, though the $( N - 2 )$ coordinates, in general,
 do not need to be mutually orthogonal.
This second criterion is shown to imply the first criterion. 
It is shown that the internal Jacobi coordinate system satisfies
 these two criteria while the internal coordinate system adopted
 by Pethick and Pitaevskii for the construction 
 of the internal one-particle density matrix does not.  
It is demonstrated that these two criteria uniquely determine
 the internal one-particle density matrix which is identical to
 that calculated with the Jacobi coordinates. 
The relevance of this work concerning $\alpha$-particle condensates 
 in nuclei, as well as bosonic atoms in traps, is pointed out.

\end{abstract}

\date{\today}
\maketitle

\section{Introduction}

The second $0^+$ state of $^{12}$C which is located near the 
 3$\alpha$ breakup threshold and is called the Hoyle 
 state~\cite{hoyle} is well known as one of the mysterious 
 $0^+$ states in light nuclei. 
The understanding of its structure 
 has been actually one of the most difficult and challenging 
 problems of nuclear structure.  
Its small excitation energy of 
 7.66 MeV is clearly not easy to explain by the shell model and,  
 in fact, even the most advanced modern shell model approach known as the 
 no-core shell model, fails by far to reproduce
 its position~\cite{nocore}. 

About 40 years ago Morinaga proposed to assign the 3$\alpha$ 
 linear-chain structure to this Hoyle state~\cite{morinaga}.  
However the observed reduced $\alpha$ decay width of this state which is 
 larger than the Wigner limit value was pointed out to be contradictory 
 to the linear-chain structure which can give at most only one third 
 of the Wigner limit value~\cite{suzhorike}.  
The large $\alpha$-decay reduced width of the Hoyle state
 in the $^8$Be$(0_1^+)$+$\alpha$ channel was reproduced by
 a full three-body calculation with a semi-microscopic $\alpha$-$\alpha$
 interaction~\cite{hori}, namely an OCM
 ( orthogonality condition model )~\cite{saito} calculation 
 for the 3$\alpha$ system. 
This 3$\alpha$ calculation contradicted the 
 3$\alpha$ chain structure of the Hoyle state and gave as the 
 dominant component of the Hoyle state the $^8$Be$(0_1^+)$ + $\alpha$ 
 structure with relative {\it S} wave between $^8$Be$(0_1^+)$ 
 and $\alpha$.  
Since $^8$Be$(0_1^+)$ consists of two $\alpha$ clusters 
 weakly coupled in relative {\it S} wave, the Hoyle state was concluded to 
 have a weakly coupled 3$\alpha$ structure in relative {\it S} waves with 
 large spatial extent, which is a gas-like structure of $\alpha$ clusters.  
A few years later, this understanding of the structure of the Hoyle 
 state was reported to be reproduced by full microscopic 3$\alpha$ 
 calculations by two groups, namely Kamimura and his 
 collaborators~\cite{kamimura} and Uegaki and his collaborators~\cite{uegaki}. 
These calculations nicely reproduced not only the 
 position of the Hoyle state but also other experimental properties 
 including inelastic electron form factor and E0 and E2 transition 
 properties. 
Other states of $^{12}$C below and around the Hoyle 
 state were also well described.  
The microscopic 3$\alpha$ model 
 treatments of $^{12}$C of Refs.~\cite{kamimura,uegaki} were 
 later extended to reaction theory in Ref.~\cite{baye}. 
Furthermore investigation by antisymmetrized molecular dynamics~\cite{enyo} and 
 that by fermionic molecular dynamics~\cite{neff} starting from a 
 realistic bare $N$-$N$ force, both of which do not assume alpha 
 clustering, have now reproduced all the salient features of the Hoyle 
 state~\cite{hoyle}. 
Let us also mention that for a product state of only a few bosons,
 the phase fluctuates.
However, we use the word 'condensate' in the same 'loose' sense
 as nuclear physicists are used to say that 'nuclei are superfluid',
 in spite of the fact that there is not a macroscopic number of
 Cooper pairs in nucleus.

Almost 30 years after the first proposal of the $^8$Be$(0_1^+)$ 
+ $\alpha$ structure for the Hoyle state, this state was reconsidered 
in a new light in Ref.~\cite{thsr} involving four members of the present 
authors. They presented the following new model wave function 
$\Phi_B (3 \alpha)$, called the THSR wave function:
\begin{eqnarray}
   \Phi_B (3 \alpha) &=& {\cal A} \{ \exp  
   [-\frac{2}{B^2} (\Vec{X}_1^2 + \Vec{X}_2^2 + \Vec{X}_3^2)] \ 
   \phi(\alpha_1)\phi(\alpha_2)\phi(\alpha_3) \} \label{eq:THSR_cm}\\ 
   &=& \exp( -\frac{6}{B^2} \Vec{\xi}_3^2 ) {\cal A} \{ 
   \exp( -\frac{4}{3B^2} \Vec{\xi}_1^2 - \frac{1}{B^2} \Vec{\xi}_2^2 ) \ 
   \phi(\alpha_1)\phi(\alpha_2)\phi(\alpha_3) \}, \label{eq:THSR_jacobi}\\ 
   \Vec{\xi}_1 &=& \Vec{X}_1 - \frac{1}{2} ( \Vec{X}_2 + \Vec{X}_3 ), \ 
   \Vec{\xi}_2 = \Vec{X}_2 - \Vec{X}_3, \  \Vec{\xi}_3 = \frac{1}{3} 
   ( \Vec{X}_1 + \Vec{X}_2 + \Vec{X}_3 ),\label{eq:jacobi_coordinates_12C}
\end{eqnarray} 
 where $\phi(\alpha_j)$ and $\Vec{X}_j$ stand for the internal wave 
 function and the center-of-mass coordinate of the $j$-th $\alpha$ cluster,  
 respectively, and ${\cal A}$ is the antisymmetrizer of the nucleons.  
As shown in Eq.~(\ref{eq:THSR_jacobi}), the THSR wave function can
 be regarded as expressing 
 the cluster structure where a $^8$Be$(0_1^+)$-like cluster 
 ${\cal A} \{ \exp( -(1/B^2) \Vec{\xi}_2^2 ) \phi(\alpha_2)\phi(\alpha_3) \}$ 
 and the $\alpha_1$ cluster couple via an {\it S}-wave inter-cluster wave 
 function $\exp( -(4/3B^2) \Vec{\xi}_1^2 )$. 
On the other hand, Equation~(\ref{eq:THSR_cm}) 
 shows that the THSR wave function represents the state where three 
 $\alpha$ clusters occupy the same single 0{\it S}-orbit 
 $\exp(-(2/B^2)\Vec{X}^2)$, namely a 3$\alpha$ condensate state which 
 is the finite size counterpart of the macroscopic $\alpha$-particle 
 condensation in infinite nuclear matter at low density~\cite{roepke}. 
What the authors of Ref.~\cite{thsr} proposed was that the 
 $^8$Be$(0_1^+)$ + $\alpha$ structure of the Hoyle state can be regarded 
 as being a 3$\alpha$ condensate state and that one can expect  
 in general the existence of $n \alpha$ condensate states in the 
 vicinity of the $n \alpha$ breakup threshold in $4n$ self-conjugate 
 nuclei~\cite{yamada04}. 
It was soon discovered~\cite{funaki} that the microscopic 
 $3 \alpha$ wave functions of both of Refs.~\cite{kamimura,uegaki}
 have overlaps of more than 95\% with a single THSR 
 wave function with a large size parameter $B$, implying small 
 overlaps between individual $\alpha$ clusters. 

This above-mentioned striking fact reported in Ref.~\cite{funaki} means 
without doubt that the Hoyle state structure has a strong relation 
with the $\alpha$ condensation physics in dilute infinite nuclear 
matter. One of the important tasks for the clarification of this 
relation is to study the magnitude of the component of the ideal 
Bose-Einstein condensation of structureless $\alpha$ particles 
which is contained in the Hoyle state wave function.  This was 
investigated by the authors of Refs.~\cite{matsumura,yamada}
 who solved the eigenvalue problem of the internal 
one-particle density matrix of the Hoyle state wave functions 
where the center-of-mass coordinate is eliminated.  As the 
internal coordinates for calculating the internal one-particle 
density matrix $\rho$ they used the Jacobi coordinates, namely 
$\Vec{\xi}_1$ and $\Vec{\xi}_2$ in Eq.~(\ref{eq:jacobi_coordinates_12C}).  
They obtained a maximum eigenvalue larger than $0.7$ for the normalized density 
 matrix $\rho$ ( Tr $\rho = 1$ ), which means that the corresponding 
 eigenfunction which is a $0S$-orbit is occupied to more than 
 $70$\% by the three $\alpha$ particles.  
This is a large percentage corroborating the almost ideal
 $\alpha$-particle condensation nature of the Hoyle state, that is
 the Hoyle state is describable to very good approximation
 by a product state of these bosons,
 each in the same $0S$ orbit~\cite{matsumura,yamada}. 

For the discussion of the Bose-Einstein condensation of a confined 
 macroscopic system, one uses in general the one-particle 
 density matrix in the laboratory frame where the coordinate 
 system consisting of individual particle coordinates is adopted. 
However, Pethick and Pitaevskii~\cite{PP} (PP) proposed to use 
 the internal one-particle density matrix by eliminating the 
 center-of-mass coordinate when one discusses the condensate 
 fraction of the system where only the center-of-mass degree of 
 freedom is excited but all relative degrees of freedom are kept 
 unchanged.  
Their proposal was to point out that the conclusion 
 of the paper by Wilkin et al.~\cite{wilkin} does not appropriately
 reflect the physics.  
Wilkin et al. discussed the lowest excitation of 
 a condensate of cold bosonic atoms with attractive interactions
 which rotates with its center of mass in a harmonic
 potential but keeps all
 the internal degrees of freedom as in its ground state
 which shows complete condensation. 
It was found that the corresponding one-body density matrix 
 in the laboratory frame has many eigenvalues of comparable 
 size, and thus the system should be characterized as a 
 fragmented condensate~\cite{wilkin}.  
On the other hand PP 
 claimed that if one uses an internal one-particle density 
 matrix by eliminating the center-of-mass coordinate it 
 should give a single eigenvalue of the order of 
 the number of particles indicating the non-fragmented 
 condensate character of the internal part of the system. 
Unfortunately PP presented only their idea and did not 
 demonstrate explicit results of the diagonalization of their 
 definition of the internal one-particle density matrix. 

In this paper, following the definition of PP, we construct the 
 internal one-particle density matrix of a many-boson system in 
 a harmonic trap and then give the explicit analytical form of 
 the eigenvalues and eigenfunctions of this density matrix. 
We will see that the eigenvalues are fragmented, which is 
 contrary to PP's initial objective. 
We will discuss that this result does not mean that
 the original idea of PP is incorrect but it means that
 the internal coordinate system which PP adopted is inadequate. 
Actually when we construct the internal 
 one-particle density matrix by using the internal Jacobi 
 coordinates, the resulting eigenvalues show complete 
 condensation.  
We will give two criteria for the choice of the adequate
 coordinate systems which will give us a unique answer for the 
 definition of the internal one-particle density matrix. 
One criterion is that in the infinite particle number limit ( $N=\infty$ )
 the internal one-particle density matrix should have the same
 eigenvalues and eigenfunctions as those of the corresponding
 ideal Bose-Einstein  condensate in the laboratory frame.  
This is in line with the general wisdom that in the thermodynamic
 limit all macroscopic quantities shall be the same,
 regardless whether considered in the laboratory frame or
 in the internal frame~\cite{blaizot-ripka}. 
The other criterion is that the coordinate used in the internal one-particle density matrix should be orthogonal to
 (or maximally independent from) the remaining $( N - 2 )$
 internal coordinates, though the $( N - 2 )$ coordinates, in general, do not need to be mutually orthogonal.  
This second criterion is shown to imply the first criterion. 
We show that the internal Jacobi coordinate system satisfies
 these two criteria, while the internal coordinate system adopted
 by PP for the construction of the internal one-particle density
 matrix does not.  
Furthermore we argue that these two criteria uniquely 
 determine the internal one-particle density matrix which is the 
 same as that calculated with the Jacobi coordinates.  
The results of this paper justify the use of the Jacobi 
coordinates in Refs.~\cite{matsumura,yamada} and, hence,  
corroborate the almost ideal $\alpha$-particle condensation nature 
of the Hoyle state.     

The present paper is organized as follows.  In Sec.~II, we 
formulate the internal one-particle density matrix with respect 
to the internal wave function of a Bose-Einstein condensate with 
finite particle number in a harmonic trap.  Then, the analytical 
form of the eigenvalues and eigenfunctions is presented for the 
internal density matrix and numerical eigenvalues are discussed.  
In Sec.~III. we propose two criteria for the choice of proper 
coordinate systems for internal one-particle density matrix. 
Finally, the summary is given in Sec.~IV. 
Appendix~\ref{app:A} serves to present the analytical solution of the eigenvalue
 problem of the density matrix and Appendix~\ref{app:B} is given for
 the explanation of the bosonic symmetry of the Jacobi-type internal
 one-particle density matrix. 
The original version of this paper was given in an article on 
 the arXiv~\cite{arxiv} and a short version of this paper 
 is reported in a letter paper~\cite{letter} with special
 attention to the cold atom community.

\section{Internal One-Particle Density Matrix}

First we consider the one-particle density matrix in the 
laboratory frame for an ideal Bose-Einstein condensate with 
$N$ spinless bosons in a harmonic potential. The result is 
trivial but instructive for studying the nature of the internal 
one-particle density matrix, as will be discussed later.  

The $N$-particle Hamiltonian in laboratory frame is presented 
as 
\begin{eqnarray}
 H = {\sum_{i=1}^{N} \frac{1}{2m}\Vec{p}_i^2} + 
   \sum_{i=1}^N \frac{1}{2}m\omega^2 \Vec{r}_i^2. 
   \label{eq:total_hamiltonian}
\end{eqnarray}
The normalized ground-state wave function of this system is expressed as 
a product of identical Gaussian single-particle wave functions, 
i.e. 
\begin{eqnarray}
\Phi(\{\Vec{r}_i\}_{i=1}^N) &=& \prod_{i=1}^{N} 
   \left( \frac{2\nu}{\pi} \right)^{3/4} 
   \exp\left(-\nu\Vec{r}_i^2\right) \\ 
  &=& \left( \frac{2\nu}{\pi} \right)^{3N/4} 
   \exp\left(-\nu \sum_{i=1}^{N} \Vec{r}_i^2 \right)
   \label{eq:total_wf}
\end{eqnarray}
where $\nu=m\omega/2\hbar$, and $\{\Vec{r}_i\}_{i=1}^N$ denotes 
the set of the coordinates $\Vec{r}_i$ ($i=1\cdots,N$). 
The one particle density matrix in the laboratory frame is 
defined as
\begin{eqnarray}
{\rho^{(1)}_{\rm Lab}(\Vec{r}, \Vec{r}')} &=& \int d\Vec{r}_2 
   d\Vec{r}_3 \cdots d\Vec{r}_N 
   \Phi^*(\Vec{r},\Vec{r}_2,\Vec{r}_3,\cdots,\Vec{r}_N) 
   \Phi(\Vec{r}',\Vec{r}_2,\Vec{r}_3,\cdots,\Vec{r}_N),\\
  &=& \left(\frac{2\nu}{\pi}\right)^{3/2} 
   \exp\left[-\nu(\Vec{r}^2+\Vec{r}'^2)\right].
   \label{eq:one-body_density}
\end{eqnarray}
It is noted that the density matrix is independent of the number 
of particles $N$ and is separable with respect to $\Vec{r}$ and 
$\Vec{r}^{\prime}$.  The separability originates from the fact 
that the Hamiltonian is separable with respect to different 
particle operators, $\Vec{r}_i$ and $\Vec{p}_i$ ($i=1,\cdots,N$), 
in Eq.~(\ref{eq:total_hamiltonian}), or equivalently from the 
fact that the wave function of Eq.~(\ref{eq:total_wf}) is 
separable with respect to different particle coordinates, 
$\Vec{r}_i$ ($i=1,\cdots,N$). 
 
The nature of the single particle orbits and their occupation 
probabilities in the relevant system can be obtained by solving 
the eigenvalue problem of the density matrix,
\begin{eqnarray}
\int \rho^{(1)}_{\rm Lab}(\Vec{r},\Vec{r}') \phi(\Vec{r}') 
   d\Vec{r}' = \lambda \phi(\Vec{r}),
\end{eqnarray}
where $\phi(\Vec{r})$ is the single particle orbit, and 
$\lambda$ is its occupation probability.  This equation can easily 
be solved, and we find that the density matrix has only one non zero 
eigenvalue $\lambda = 1$ with one eigenfunction, namely, the 
zero-node $S$-wave Gaussian 
$\phi(\Vec{r})=(2\nu/\pi)^{3/4}\exp(-\nu\Vec{r}^2)$ 
(or $0S$ harmonic oscillator wave function $\phi_{000}(\Vec{r},\nu)$, 
which will be defined later) with 100~\% occupancy ($\lambda=1$). 
This means that all particles are condensed in that single orbit, 
i.e. an ideal Bose-Einstein condensation is realized in the 
laboratory frame.  This feature is independent of the number of 
particles $N$.

Next we consider the internal one-particle density matrix for 
 the $N$-particle Bose-Einstein condensation in a harmonic trap 
 described by the wave function Eq.~(\ref{eq:total_wf})  with the 
 total Hamiltonian Eq.~(\ref{eq:total_hamiltonian}).  
{\it Internal} means that the density is free from the center-of-mass
 coordinate of the system, as it must be the case, when
 dealing with, e.g., selfbound systems.  
In the present paper, two kinds of internal 
 coordinate sets are introduced;  1)~coordinates with respect to 
 the center of mass of the total system and 2)~Jacobi coordinates. 
The former set was 
 first considered by Pethick and Pitaevskii~\cite{PP} to define 
 the internal one-particle density matrix.  We call it 
 Pethick-Pitaevskii-type (PP-type) internal one-particle density 
 matrix in the present paper. For the latter set, we call it 
 Jacobi-type density matrix.

\subsection{Pethick-Pitaevskii-type internal one-particle density matrix}

As already mentioned in the introduction, Pethick and Pitaevskii proposed to
 consider the internal single particle density matrix for a Bose condensed
 system when only the center-of-mass but no internal degree of freedom
 is excited.
In order to define an internal one-particle density matrix, Pethick 
 and Pitaevskii adopted internal coordinates defined with respect to 
 the center of mass of the total system~\cite{PP}. 
The center-of-mass coordinate $\Vec{R}$ and the coordinate $\Vec{q}_i$
 of particle $i$ relative to the center of mass are given, respectively, by
\begin{eqnarray}
\Vec{R}=\frac{1}{N}\sum_{i=1}^N \Vec{r}_i, \hspace*{5mm}
   \Vec{q}_i=\Vec{r}_i-\Vec{R}. \label{eq:coordinate_cm_rel}
\end{eqnarray}
Since $\sum_{i=1}^N \Vec{q}_i = 0$,  only $N-1$ of the $\Vec{q}_i$ are 
 independent.  
Hereafter, as the independent internal coordinates, we adopt 
 $\Vec{q}_i$ ($i=1,\cdots,N-1$).  
We introduce the conjugate momenta $\Vec{\pi}_i$ ($i=1,\cdots,N-1$)
 and $\Vec{P}$ for the coordinates $\Vec{q}_i$ ($i=1,\cdots,N-1$) and $\Vec{R}$, respectively.  
Then, the total Hamiltonian in Eq.~(\ref{eq:total_hamiltonian}) is rewritten as
\begin{eqnarray}
&&H = H_{\rm int} + H_{\rm cm},\label{eq:PP_Hamiltonian} \\
&&H_{\rm int}=\frac{1}{2m} \left[ {\left(\frac{N-1}{N}\right) 
  \sum_{i=1}^{N-1} \Vec{\pi}_{i}^{2}} - {\frac{2}{N} \sum_{i<i'=1}^{N-1} 
  \Vec{\pi}_i\cdot\Vec{\pi}_{i'}} \right] 
  + m\omega^2 \left[ \sum_{i=1}^{N-1}{\Vec{q}_i}^2 + 
  \sum_{i<i'=1}^{N-1}\Vec{q}_i\cdot\Vec{q}_{i'}\right], 
  \label{eq:PP_internal_Hamiltonian} \\
&&H_{\rm cm}=\frac{1}{2Nm}\Vec{P}^2 + \frac{1}{2}Nm\omega^2\Vec{R}^2,
\end{eqnarray}
where $H_{\rm int}$ and $H_{\rm cm}$ denote the internal and 
center-of-mass Hamiltonians, respectively.
It should be mentioned that the above Hamiltonian includes the cross terms,
 $\Vec{q}_i\cdot\Vec{q}_{i'}$ and $\Vec{\pi}_i\cdot\Vec{\pi}_{i'}$, which,
 in general, appear so far as one takes non-orthonormal coordinate
 systems~\cite{clobt1,clobt2,clobt3,clobt4}.

By using the relation 
\begin{eqnarray}
 \sum_{i=1}^N \Vec{r}_i^2 = N \Vec{R}^2 + \sum_{i,i'=1}^{N-1} 
  (\delta_{i,i'} + 1) \Vec{q}_i \cdot \Vec{q}_{i'}, 
  \label{eq:PPquadra}
\end{eqnarray}
the total wave function in Eq.~(\ref{eq:total_wf}) is expressed as
\begin{eqnarray}
&&\Phi(\{\Vec{r}_i\}_{i=1}^{N}) = \frac{1}{N^{3/2}} \times  
  \Phi_{\rm int}(\{\Vec{q}_i\}_{i=1}^{N-1}) \Phi_{\rm cm}(\Vec{R}), 
  \label{eq:wf_cm_rel}\\
&&\Phi_{\rm int}(\{\Vec{q}_i\}_{i=1}^{N-1}) = 
  \left( \frac{N\times (2\nu)^{N-1}}{\pi^{N-1}} \right)^{3/4} 
  \exp\left[-\sum_{i,i'=1}^{N-1} (\delta_{ii'}+1) \nu \Vec{q}_i 
  \cdot \Vec{q}_{i'}\right],
  \label{eq:PP_wf_internal} \\
&&\Phi_{\rm cm}(\Vec{R})=\left( \frac{2N\nu}{\pi} \right)^{3/4} 
  \exp(-N\nu\Vec{R}^2),
  \label{eq:PP_wf_cm}
\end{eqnarray}
where $\Phi_{\rm int}$ and $\Phi_{\rm cm}$ denote the internal and 
center-of-mass wave functions, respectively. The separability of 
$\Phi_{\rm int}$ and $\Phi_{\rm cm}$ comes from the fact that the 
total Hamiltonian is separable with respect to the internal and 
center-of-mass parts [see Eq.~(\ref{eq:PP_Hamiltonian})] or directly 
from Eq.~(\ref{eq:PPquadra}) inserted 
into Eq.~(\ref{eq:total_wf}).  The factor $1/N^{3/2}$ in 
Eq.~(\ref{eq:wf_cm_rel}) drops out when taking into account the 
Jacobian $\partial (\{ \Vec{r}_i \}_{i=1}^{N})/\partial 
(\{ \Vec{q}_i \}_{i=1}^{N-1},\Vec{R})=N^3$, coming from the 
coordinate transformation from the space-fixed system to the 
center-of-mass system. It should be noted that both wave functions 
$\Phi_{\rm int}$ and $\Phi_{\rm cm}$ are normalized and satisfy 
$H_{\rm int}\Phi_{\rm int}=(3/2)(N-1)\hbar\omega\Phi_{\rm int}$ 
and $H_{\rm cm}\Phi_{\rm cm}=(3/2)\hbar\omega\Phi_{\rm cm}$, 
respectively.

According to Pethick and Pitaevskii~\cite{PP}, the internal 
one-particle density matrix is defined as,
\begin{eqnarray}
&&\rho_{\rm int, PP}^{(1)}(\Vec{q},\Vec{q}^\prime) = 
  \int d\Vec{q}_2 \cdots d\Vec{q}_{N-1}~\rho_{\rm int, PP}
  (\Vec{q},\Vec{q}_2,\cdots,\Vec{q}_{N-1};\Vec{q}^\prime,
  \Vec{q}_2,\cdots,\Vec{q}_{N-1}), 
  \label{eq:one_body_den_PP} \\
&&\rho_{\rm int, PP}(\{\Vec{q}_i\}_{i=1}^{N-1}, 
  \{\Vec{q}_i^\prime\}_{i=1}^{N-1})
  =N^3 \int d\Vec{R}~\Phi^*(\{\Vec{q}_i+\Vec{R}\}_{i=1}^{N}) 
  \Phi(\{\Vec{q}_i^\prime+\Vec{R}\}_{i=1}^{N}),
  \label{eq:internal_density_PP}
\end{eqnarray}
where the Jacobian factor $N^3$ is inserted in the above 
equation, which is missing in Ref.~\cite{PP}. Using the wave 
function $\Phi(\{\Vec{r}_i\}_{i=1}^{N})$ in 
Eq.~(\ref{eq:total_wf}), the internal one-particle density 
matrix (\ref{eq:one_body_den_PP}) is expressed as
\begin{eqnarray}
\rho_{\rm int,PP}^{(1)}(\Vec{q},\Vec{q}^\prime) = 
  \left( \frac{N}{N-1} \right)^{3/2} 
  \left( \frac{2\nu}{\pi} \right)^{3/2} 
  \exp\left[-\frac{3N-2}{2(N-1)}\nu(\Vec{q}^2 + 
  {\Vec{q}^\prime}^2) 
  + \frac{N-2}{N-1}\nu\Vec{q}\cdot\Vec{q}^\prime \right].
  \label{eq:one_body_den_int} 
\end{eqnarray}
It is noted that this density matrix depends on the number 
of particles $N$ and contains the cross term 
$\Vec{q}\cdot\Vec{q}^\prime$.  The origin of the cross term 
comes from the nonseparability of the coordinate $\Vec{q}_1$ 
from the other $(N-2)$ coordinates $\Vec{q}_i$ ($i=2,\cdots,N-1$) 
in the internal wave function of Eq.~(\ref{eq:PP_wf_internal}), or 
equivalently from the fact that the internal Hamiltonian of 
Eq.~(\ref{eq:PP_internal_Hamiltonian}) is non-separable between 
the first particle operators $\Vec{\pi}_1$ and $\Vec{q}_1$ and the 
other $(N-2)$ particle operators $\Vec{\pi}_i$ and $\Vec{q}_i$ 
($i=2,\cdots,N-1$).  The result of Eq.~(\ref{eq:one_body_den_int}) 
can also be verified with the use of the internal density 
matrix in Eq.~(\ref{eq:internal_density_PP}) presented as 
follows: 
\begin{eqnarray}
\rho_{\rm int, PP}(\{\Vec{q}_i\}_{i=1}^{N-1}, 
  \{\Vec{q}_i^\prime\}_{i=1}^{N-1}) = \Phi_{\rm int}^*
  (\{\Vec{q}_i\}_{i=1}^{N-1}) \Phi_{\rm int}
  (\{\Vec{q}^\prime_i\}_{i=1}^{N-1}),
  \label{eq:PP_internal_density_int}
\end{eqnarray}
where $\Phi_{\rm int}^*(\{\Vec{q}_i\}_{i=1}^{N-1})$ is the 
internal wave function defined in Eq.~(\ref{eq:PP_wf_internal}).

Let us discuss the nature of the internal one-particle density 
 $\rho_{\rm int}^{(1)}(\Vec{q},\Vec{q}^\prime)$. 
First we study the single-particle orbits and their eigenvalues 
 obtained by solving the eigenvalue problem for the density matrix,
\begin{eqnarray}
\int \rho_{\rm int,PP}^{(1)}(\Vec{q},\Vec{q}^\prime) 
  \phi(\Vec{q}) d\Vec{q}^\prime = \lambda \phi(\Vec{q}). 
\end{eqnarray}
This equation can be solved analytically. 
There are several ways to solve it. 
In Appendix A we give one method. 
We can also analyze the description given in Ref.~\cite{gajda}.  
The single-particle orbits 
 $\phi$ are expressed by the harmonic oscillator wave functions 
 $\phi_{nLM}(\Vec{q},\beta_N)$
 with the orbital angular momentum $L$, 
 magnetic quantum number $M$ and harmonic oscillator quanta
 $Q=2n+L$ $(n=0,1,2,\cdots)$,
\begin{eqnarray}
&&\phi_{nLM}(\Vec{q},\beta_N) = \left[\frac{(2\beta_N)^{L+3/2}2^{n+L+2}n!}
  {\sqrt{\pi}(2n+2L+1)!!} \right]^{1/2} q^L L_n^{(L+1/2)}(2\beta_N q^2) 
  {\exp (-\beta_N q^2)} Y_{LM}(\hat{\Vec{q}}),  \label{eq:harmosci} \\
&&\beta_N=\sqrt{\frac{2N}{N-1}}\nu.
\end{eqnarray}
The eigenvalues or occupation probabilities $\lambda$ are given as
\begin{eqnarray}
\lambda^{(LM)}_{n,N} = 
  \left[ \frac{4N}{3N-2+2\sqrt{2N(N-1)}} \right]^{3/2}
  \left[ \frac{N-2}{3N-2+2\sqrt{2N(N-1)}} \right]^{2n+L},
\end{eqnarray}
and satisfy the following completeness relation,
\begin{eqnarray}
\sum_{L=0}^{\infty} \sum_{M=-L}^{L} \sum_{n=0}^{\infty} 
  \lambda^{(LM)}_{n,N} =1.
\end{eqnarray}
Then, the internal one-particle density matrix for the $N$ particles 
 in Eq.~(\ref{eq:one_body_den_int}) can be presented in terms of
 the wave functions (\ref{eq:harmosci})
\begin{eqnarray}
\rho_{\rm int,PP}^{(1)}(\Vec{q},\Vec{q}^\prime) 
  = \sum_{L=0}^{\infty} \sum_{M=-L}^{L}\sum_{n=0}^{\infty}\  
    {| \phi_{nLM}(\Vec{q},\beta_N) \rangle}\ \lambda^{(LM)}_{n,N}\ 
    {\langle  \phi_{nLM}(\Vec{q}^{\prime},\beta_N)|}.
\end{eqnarray}
The occupation probability with respect to the partial wave with quantum 
number $L$ is defined as
\begin{eqnarray}
\Lambda^{(L)}_N = \sum_{M=-L}^{L} \sum_{n=0}^{\infty} 
  \lambda^{(LM)}_{n,N}.
  \label{eq:PP_occupation_probablity_N_L}
\end{eqnarray}
In the macroscopic limit ($N\rightarrow \infty$), the internal one-particle 
density matrix $\rho_{{\rm int,PP},N=\infty}^{(1)}$,  its eigenfunctions 
$\phi_{{\rm int,PP},N=\infty}$, and eigenvalues 
$\lambda^{(LM)}_{n,N=\infty}$ are given by
\begin{eqnarray}
&&\rho_{{\rm int,PP},N=\infty}^{(1)}(\Vec{q},\Vec{q}^\prime) = 
  \left( \frac{2\nu}{\pi} \right)^{3/2} 
  \exp\left[-\frac{3}{2}\nu(\Vec{q}^2+{\Vec{q}^\prime}^2) + 
  \nu\Vec{q}\cdot\Vec{q}^\prime \right], 
  \label{eq:one_body_den_int_infinite} \\
&&\phi_{{\rm int,PP},N=\infty}(\Vec{q})=\phi_{nLM}(\Vec{q},\sqrt{2}\nu),
  \label{eq:one_body_den_int_infinite_eigenfunction} \\
&&\lambda^{(LM)}_{n,N=\infty}=2^3 (3-2\sqrt{2})^{2n+L+3/2},
  \label{eq:one_body_den_int_infinite_eigenvalue} \\
&&\Lambda^{(L)}_{N=\infty} = \sum_{M=-L}^{L} 
  \sum_{n=0}^{\infty}\lambda^{(LM)}_{n,N=\infty} = 
  (2L+1)(2-\sqrt{2})(3-2\sqrt{2})^L.
  \label{eq:eq:one_body_den_int_infinite_probability}
\end{eqnarray}
We remark that the summed eigenvalues $\Lambda^{(L)}_{N=\infty}$ still 
depend on the angular momentum $L$.

The eigenvalue or occupation probability of the PP-type internal
 one-particle density matrix, $\Lambda_N^{(L)}$ of
 Eq.~(\ref{eq:PP_occupation_probablity_N_L}) depends on
 the orbital angular momentum $L$, and the particle number $N$.
In the case of $N=3$, we obtain the occupation 
 probabilities, $\Lambda_3^{(S)}=0.804$ for $S$ wave, 
 $\Lambda_3^{(P)}=0.173$ for $P$, $\Lambda_3^{(D)}=0.021$ for $D$, 
 $\Lambda_3^{(F)}=0.002$ for $F$, and so on.  
Increasing the particle number, the $S$-wave occupation probability
 is decreasing, while the higher partial-wave ones are increasing
 (see Fig.~1 in Refs.~\cite{arxiv,letter}). 
These results show that the PP-type one-particle density matrix
 leads to a fragmented condensate. 
The reason why the PP-type internal density matrix shows the fragmented 
 condensate is due to the existence of the cross term 
 $\Vec{q}\cdot\Vec{q}^{\prime}$ in Eq.~(\ref{eq:one_body_den_int}).

\subsection{Jacobi-type internal one-particle density matrix}

For the $N$-particle system, we define the $(N-1)$ internal Jacobi 
coordinates $\{ \Vec{\xi}_i, (i=1, \cdots, N-1) \}$ and the 
center-of-mass coordinate $\Vec{R}$ as follows:
\begin{eqnarray}
&&\Vec{\xi}_i = \Vec{r}_i -{\frac{1}{N-i}\sum_{k=i+1}^{N}\Vec{r}_k}, 
  \hspace*{5mm}{\rm for}~~{i=1, 2, \cdots, {N-1}}
  \label{eq:Jacobi_coordinates}\\
&&\Vec{R} = {\frac{1}{N}\sum_{i=1}^{N} \Vec{r}_i},
  \label{eq:Jacobi_coordinates_cm}
\end{eqnarray}
where $\Vec{\xi}_1$ denotes the relative coordinate between the first 
particle and the remaining $(N-1)$ particles, and 
other Jacobi coordinates are self-evident.  Then, the $N$-particle 
Hamiltonian in Eq.~(\ref{eq:total_hamiltonian}) can be separated into 
the internal and center-of-mass Hamiltonian,
\begin{eqnarray}
&&H = H_{\rm int} + H_{\rm cm},
  \label{eq:Jacobi_Hamiltonian} \\
&&H_{\rm int}= \sum_{i=1}^{N-1} \{ \frac{1}{2\mu_i} \Vec{\pi}_i^2 
  + \frac{1}{2} \mu_i {\omega}^2 \Vec{\xi}_i^2 \}, 
  \hspace*{5mm} \mu_i = \frac{N-i}{N-i+1} m,
  \label{eq:Jacobi_internal_Hamiltonian} \\
&&H_{\rm cm}=\frac{1}{2Nm}\Vec{P}^2 + \frac{1}{2}Nm\omega^2\Vec{R}^2,
  \label{eq:Jacobi_cm_Hamiltonian}
\end{eqnarray}
where $\Vec{\pi}_i$ and $\Vec{P}$ denote the conjugate momenta 
corresponding to the coordinates $\Vec{\xi}_i$ and $\Vec{R}$, 
respectively.

Since the total Hamiltonian is a sum of decoupled $N$ harmonic oscillator 
Hamiltonians ( for $\Vec{R}$ and $(N-1)$ internal Jacobi coordinates ), the 
total wave function in Eq.~(\ref{eq:total_wf}) is expressed as 
\begin{eqnarray}
\Phi(\{\Vec{r}_i\}_{i=1}^{N}) &=& \Phi_{\rm int}(\{\Vec{\xi}_i\}_{i=1}^{N-1}) 
  \Phi_{\rm cm}(\Vec{R}),
  \label{eq:Jacobi_wf_cm_rel} \\
\Phi_{\rm int}(\{\Vec{\xi}_i\}_{i=1}^{N-1}) &=& \prod_{i=1}^{N-1} 
  \left( \frac{2\nu_i}{\pi} \right)^{3/4} \exp\left(-\nu_i\Vec{\xi}_{i}^2 
    \right) \\ 
  &=& \left( \frac{(2\nu)^{N-1}}{N\times \pi^{N-1}} \right)^{3/4} 
    \exp \{ -\sum_{i=1}^{N-1} \nu_i \Vec{\xi}_i^2 \}, \hspace*{5mm} 
    \nu_i=\frac{N-i}{N-i+1}\nu,
  \label{eq:Jacobi_wf_internal} \\
\Phi_{\rm cm}(\Vec{R}) &=& \left( \frac{2N\nu}{\pi} \right)^{3/4} 
  \exp(-N\nu\Vec{R}^2).
\end{eqnarray}
This expression of $\Phi(\{\Vec{r}_i\}_{i=1}^{N})$ can also be derived by 
inserting the relation 
\begin{eqnarray}
 \sum_{i=1}^N \Vec{r}_i^2 = N \Vec{R}^2 + 
   \sum_{i=1}^{N-1} \frac{\nu_i}{\nu} \Vec{\xi}_i^2, 
   \label{eq:Jacobi_quadra}
\end{eqnarray}
into the original expression given in  Eq.~(\ref{eq:total_wf}).  The total 
wave function is the product of the internal and center-of-mass wave 
functions, $\Phi_{\rm int}$ and $\Phi_{\rm cm}$, which are normalized and 
satisfy the relations, $H_{\rm int}\Phi_{\rm int}=(3/2)(N-1)\hbar\omega
\Phi_{\rm int}$ and $H_{\rm cm}\Phi_{\rm cm}=(3/2)\hbar\omega\Phi_{\rm cm}$. 
It is noted that the internal wave function is given as a product of 
harmonic oscillator wave functions $\phi_{000}(\Vec{\xi}_i,\nu_i)$ 
$(i=1,\cdots,N-1)$.

The Jacobi-type internal density matrix writes
\begin{eqnarray}
\rho_{\rm int, J}(\{\Vec{\xi}_i\}_{i=1}^{N-1}, 
  \{\Vec{\xi}_i^\prime\}_{i=1}^{N-1}) = \Phi_{\rm int}^* 
  (\{\Vec{\xi}_i\}_{i=1}^{N-1}) \Phi_{\rm int} 
  (\{\Vec{\xi}^\prime_i\}_{i=1}^{N-1}).
\end{eqnarray}
Then, the Jacobi-type one-particle density matrix is defined with respect to 
$\Vec{\xi}_1$ and $\Vec{\xi}_1^{\prime}$ as
\begin{eqnarray}
\rho_{\rm int,J}^{(1)}(\Vec{\xi},\Vec{\xi}^\prime) &=& 
  \int d\Vec{\xi}_2 \cdots d\Vec{\xi}_{N-1}~\rho_{\rm int, J}(\Vec{\xi},
   \Vec{\xi}_2,\cdots,\Vec{\xi}_{N-1};\Vec{\xi}^\prime,\Vec{\xi}_2,\cdots,
   \Vec{\xi}_{N-1}),\label{eq:Jacobi_one_body_den_int_general} \\ 
  &=& \left( \frac{N-1}{N} \right)^{3/2} \left( \frac{2\nu}{\pi} \right)^{3/2} 
   \exp\left[-\frac{N-1}{N}\nu(\Vec{\xi}^2+{\Vec{\xi}^\prime}^2)\right].
  \label{eq:Jacobi_one_body_den_int} 
\end{eqnarray}
This choice of the coordinate $\Vec{\xi}_1$ for the internal density matrix 
 is natural, because the single particle orbit should be defined with respect 
 to the relative coordinate between one particle and the other remaining 
 $N-1$ particles in the Jacobi coordinate system.  
The compatibility between the bosonic symmetry of the system
 and the above definition of the internal 
 one-particle density matrix, where it may seem that one special coordinate
 is singled out, is explained in Appendix~\ref{app:B}.  
The eigenvalue equation of the one-particle density matrix can easily be solved
 analytically. 
We find that the density matrix has only one non-zero eigenvalue $\lambda=1$ 
 with corresponding eigenfunction $\phi_{{\rm int,J},N}$ which is the $0S$ 
 harmonic oscillator wave function with 100~\% occupancy ($\lambda=1$),
\begin{eqnarray}
&&\phi_{{\rm int,J},N}(\Vec{\xi})=\delta_{L0}\delta_{M0} \phi_{000}
  (\Vec{\xi},(N-1)\nu/N),
  \label{eq:Jacobi_single_orbit}\\
&&\lambda^{(LM)}_{N}=\delta_{L0}\delta_{M0},\\
&&\Lambda^{(L)}_{N}=\sum_{M=-L}^{L} \lambda^{(LM)}=\delta_{L0}.
  \label{eq:Jacobi_occupation_prob}
\end{eqnarray}
This means that all particles are condensed in the single $0S$ particle state, 
 although the size parameter in the state [Eq.~(\ref{eq:Jacobi_single_orbit})] 
 depends on $N$ and is slightly different from that in the eigenfunction 
 $\phi_{000}(\Vec{r},\nu)$ in laboratory frame, discussed in the beginning of 
 Sec.~II. 
In the macroscopic limit ($N\rightarrow \infty$), the internal 
 one-particle density matrix $\rho_{{\rm int,PP},N=\infty}^{(1)}$, 
 its eigenfunction $\phi_{{\rm int,PP},N=\infty}$, and eigenvalues 
 $\lambda^{(LM)}_{n,N=\infty}$ are given by
\begin{eqnarray}
&&\rho_{{\rm int},J,N=\infty}^{(1)}(\Vec{\xi},\Vec{\xi}^\prime) = 
  \left( \frac{2\nu}{\pi} \right)^{3/2} 
  \exp\left[-\nu(\Vec{\xi}^2+{\Vec{\xi}^\prime}^2)\right], 
  \label{eq:Jacobi_one_body_den_int_infinite} \\
&&\phi_{{\rm int,J},N=\infty}(\Vec{\xi}) = \delta_{L0}
  \delta_{M0} \phi_{000}(\Vec{\xi},\nu),
  \label{eq:Jacobi_one_body_den_int_infinite_wf}\\
&&\lambda^{(LM)}_{N=\infty}=\delta_{L0}\delta_{M0},
  \label{eq:Jacobi_one_body_den_int_infinite_eigenvalue}\\
&&\Lambda^{(L)}_{N=\infty}=\sum_{M=-L}^{L}\lambda^{(LM)}_{N=\infty}
  =\delta_{L0} .
\end{eqnarray}

The Jacobi-type one-particle density matrix for finite particle 
 number is separable with respect to the coordinates $\Vec{\xi}$ 
 and $\Vec{\xi}^{\prime}$. 
This leads to the only one non-zero 
 eigenvalue $\lambda = 1$ which is the same as for an ideal 
 Bose-Einstein condensate system in the laboratory system. 
Also the internal density matrix (\ref{eq:Jacobi_one_body_den_int_infinite}) becomes 
 identical to the one in the laboratory system, Eq.~(8), for $N=\infty$.  
The separability of the internal density matrix comes from the 
 separability of the internal wave function in 
 Eq.~(\ref{eq:Jacobi_wf_internal}) with respect to different 
 Jacobi coordinates, which originates from the separability 
 of the internal Hamiltonian in 
 Eq.~(\ref{eq:Jacobi_internal_Hamiltonian}) with respect to 
 different Jacobi coordinates.  
This feature has it origin in the 
 fact that the Jacobi coordinates form an orthogonal coordinate 
 system.

\section{Criterion for the Choice of Adequate Internal Coordinates}

\subsection{Convergence to Bose-Einstein Condensation in the Macroscopic Limit}
\label{subsec:convergence_BEC_macro}

In the previous section we learned that the outcome of the 
diagonalization of the internal density matrix depends on the 
choice of the internal coordinates. This is a serious problem for 
treating condensates in internal self-bound systems such as 
$\alpha$-particle condensates in nuclear systems or small droplets of 
superfluid $^4$He, because only internal degrees of freedom are relevant 
in these systems.

In order to overcome the difficulty, we gave a criterion for the choice 
 of the internal coordinates:~In the macroscopic limit 
 ($N\rightarrow \infty$) the internal density matrix should have the same 
 eigenvalues and eigenfunctions as those of the ideal Bose-Einstein 
 condensate in the laboratory frame. 
This is a very physical boundary condition. 
For understanding this physical condition, consideration of 
 the following situation in the macroscopic limit may be helpful:~In
 the laboratory frame, the center-of-mass motion of the present 
 system is described by the wave function $\Phi_{\rm cm}(\Vec{R})$ of 
 Eq.~(\ref{eq:PP_wf_cm}).  
In the macroscopic limit, the center-of-mass 
 coordinate should be at the coordinate origin of the laboratory frame, 
 because the probability of finding the center-of-mass coordinate at 
 position $\Vec{R}$ is given by $|\Phi_{\rm cm}(\Vec{R})|^2 = 
 (2N \nu/\pi)^{3/2} \exp( -2N \nu\Vec{R}^2 )$ which becomes a 
 delta function $\delta (\Vec{R})$ in the limit of $N\rightarrow \infty$. 
Thus the internal coordinate $\Vec{q}_i$ should have the same meaning 
 as the position of the $i$-th particle coordinate $\Vec{r}_i$ in the 
 laboratory frame. 
Also the first Jacobi internal coordinate 
 $\Vec{\xi}_1$ should have the same meaning as the position of the 
 first particle coordinate $\Vec{r}_1$ because $\Vec{\xi}_1 = 
 (N/(N-1)) \Vec{q}_1 \rightarrow \Vec{r}_1$ for $N\rightarrow \infty$. 

As shown in Eqs.~(\ref{eq:one_body_den_int_infinite_eigenfunction}) and 
 (\ref{eq:one_body_den_int_infinite_eigenvalue}), the PP-type internal 
 density matrix in the macroscopic limit does not satisfy the condition,
 while the density matrix of the Jacobi-type fulfills this condition
 as given in Eqs.~(\ref{eq:Jacobi_one_body_den_int_infinite_wf}) and 
 (\ref{eq:Jacobi_one_body_den_int_infinite_eigenvalue}).
As already mentioned, 
 the reason why the PP-type internal density matrix does not satisfy the 
 physical boundary condition is due to the fact that it exhibits a 
 nonvanishing cross term or correlation term $\Vec{q}\cdot\Vec{q}^{\prime}$ 
 in the case of $N=\infty$, originating from the pseudo two-body interaction 
 terms in the internal Hamiltonian.   
On the other hand, in the Jacobi-type 
 internal density matrix, the pseudo two-body interaction terms disappear. 

These results mean that one should take internal coordinates which do 
not produce any correlation in the internal one-particle density matrix 
in the macroscopic limit. Otherwise, unphysical situations occur like for
the PP-type internal density matrix. One choice fulfilling the physical 
condition is the one of the internal Jacobi coordinates. Of course, there 
are many sets of internal coordinates which satisfy the physical 
condition. For example, they may be those satisfying the following two 
conditions:~(1)~The relative coordinate $\Vec{\xi}$ between one particle 
and the other remaining particles should be used as the coordinate of the 
internal one-particle density matrix $\rho_{\rm int}^{(1)}$, and (2)~the 
internal one-particle density matrix be separable with respect to 
$\Vec{\xi}$ and $\Vec{\xi}^{\prime}$ in the macroscopic limit, 
$\rho_{\rm int}^{(1)}(\Vec{\xi},\Vec{\xi}^{\prime})\rightarrow 
(2\nu/\pi)^{3/2}\exp[-\nu(\Vec{\xi}^2+{\Vec{\xi}^{\prime}}^2)]$ for 
$N\rightarrow \infty$. 
Among the coordinates satisfying the two conditions, Jacobi coordinates
 are very convenient and useful to describe the internal 
 Hamiltonian and thus the internal one-particle density matrix.
We will see in the next section that this is in fact
 the only choice.

\subsection{Orthogonality of the Coordinate $\Vec{\xi}_1$ with respect to
 the $(N-2)$ Internal Coordinates}

In the previous section, a criterion was discussed for the choice of
 the adequate coordinate systems of the internal density matrix.
Here we present the other criterion that the coordinate of the
 internal one-body density matrix, $\Vec{\xi}_1$, should be orthogonal to
 the remaining $(N-2)$ internal coordinates, where the $( N - 2 )$ coordinates
 do not need to be mutually orthogonal.
This is called ``maximal independence of the coordinate $\Vec{\xi}_1$ from 
 the rest of $(N-2)$ internal coordinates''.

The coordinate of the internal one-particle density matrix should be 
 either the one particle coordinate measured from the total center-of-mass 
 or the one particle coordinate measured from the center-of-mass of the 
 other $(N-1)$ particles. 
The PP-type internal one-particle density matrix 
 adopts the former type coordinate $\Vec{q}_1$ while the Jacobi-type 
 internal one-particle density matrix adopts the latter type coordinate 
 $\Vec{\xi}_1$, coinciding in the macroscopic number.
These two types of coordinates are essentially the same 
 since they are related by $\Vec{q}_1 = ((N-1)/N) \Vec{\xi}_1$.  
In constructing the internal one-particle density matrix, we integrate 
 the internal density matrix over the other $(N-2)$ internal coordinates. 
In order for the internal one-particle density matrix to have maximum 
 information on the one-particle degree-of-freedom of the system, these 
 $(N-2)$ internal coordinates should be maximally independent from the 
 coordinate $\Vec{q}_1$ or $\Vec{\xi}_1$.  
Since this requirement seems
 natural, we will study in this subsection its consequence. 
We will see below that this requirement implies the criterion proposed 
 in the preceding subsection. 

The requirement that the coordinate $\Vec{q}_1$ or $\Vec{\xi}_1$ is 
 maximally independent from the other $(N-2)$ internal coordinates means 
 mathematically that coordinate $\Vec{q}_1$ or $\Vec{\xi}_1$  
 should be orthogonal to the other $(N-2)$ internal coordinates. 
The Jacobi coordinates are just such coordinates, since they constitute 
 an orthogonal coordinate system.  
On the other hand the internal 
 coordinates $\{ \Vec{q}_i, \ i= 1 \sim (N-1) \}$ do not satisfy this 
 requirement since they constitute a non-orthogonal coordinate system.  
For the sake of self-containedness we recall here the meaning of 
 orthogonality between coordinates.   
Two coordinates $\Vec{\beta}$ and 
 $\Vec{\gamma}$ are defined to be mutually orthogonal when their 
 expansion coefficients of linear combination with respect to the 
 $N$ particle coordinates $\{ \Vec{r}_i,\ (i=1 \sim N) \}$, 
 $\{ C_i(\Vec{\beta}),\ (i=1 \sim N) \}$ and $\{ C_i(\Vec{\gamma}),\ (i=1 \sim N) \}$, 
\begin{eqnarray}
 \Vec{\beta} = \sum_{i=1}^N C_i(\Vec{\beta}) \Vec{r}_i, \ \ 
 \Vec{\gamma} = \sum_{i=1}^N C_i(\Vec{\gamma}) \Vec{r}_i,
\end{eqnarray}
are mutually orthogonal, $\sum_{i=1}^N C_i(\Vec{\beta}) C_i(\Vec{\gamma}) = 0$.  
This definition can be stated as follows. 
To any coordinate $\Vec{\delta}$ we associate an $N$-dimensional 
 number vector $\Vec{C}(\Vec{\delta})$ like $\Vec{C}(\Vec{\beta}) = 
 \{ C_i(\Vec{\beta}),\ (i=1 \sim N) \}$ for $\Vec{\beta}$ and 
 $\Vec{C}(\Vec{\gamma}) = \{ C_i(\Vec{\gamma}),\ (i=1 \sim N) \}$ 
 for $\Vec{\gamma}$. 
If $\Vec{\beta} \ne \Vec{\gamma}$, 
 $\Vec{C}(\Vec{\beta}) \ne \Vec{C}(\Vec{\gamma})$.  $\Vec{\beta}$ and 
 $\Vec{\gamma}$ are said to be orthogonal when 
 $\Vec{C}(\Vec{\beta}) \cdot \Vec{C}(\Vec{\gamma}) = 0$. 
Needless to say, the total center-of-mass coordinate $\Vec{R}$ is 
 orthogonal to any kind of internal coordinates $\Vec{\delta}_{\rm int}$,
 i.e.~$\Vec{C}(\Vec{R}) \cdot \Vec{C}(\Vec{\delta}_{\rm int}) = 0$.

Let an internal coordinate system 
 $\{ \Vec{\eta}_i, \ (i= 1 \sim (N-1)) \}$ be such a system satisfying 
 the above-mentioned requirement. 
The coordinate $\Vec{\eta}_1$ is either 
 $\Vec{q}_1$ or $\Vec{\xi}_1$ and hence we here fix it as $\Vec{\eta}_1 
 = \Vec{\xi}_1$ . The orthogonality of $\Vec{\xi}_1$ to the other $(N-2)$ 
 coordinates $\{ \Vec{\eta}_i, \ (i= 2 \sim (N-1)) \}$ means 
 $\Vec{C}(\Vec{\xi}_1) \cdot \Vec{C}(\Vec{\eta}_i) = 0, \  
 (i= 2 \sim (N-1))$. 
Since the internal Jacobi coordinate system $\{ \Vec{\xi}_i, \ (i= 1 \sim (N-1)) \}$
 is an orthogonal coordinate system, there the relations 
 $\Vec{C}(\Vec{\xi}_1) \cdot \Vec{C}(\Vec{\xi}_i) = 0, \  (i= 2 \sim (N-1))$ holds.  
Therefore the subspace spanned by the $N$-dimensional vectors
 $\{ \Vec{C}(\Vec{\eta}_i), \ (i= 2 \sim (N-1)) \}$ is identical
 to the subspace spanned by $N$-dimensional vectors
 $\{ \Vec{C}(\Vec{\xi}_i), \ (i= 2 \sim (N-1)) \}$. 
We, therefore, have 
\begin{eqnarray}
 \Vec{C}(\Vec{\xi}_i) = \sum_{j=2}^{N-1} d_{ij} \Vec{C}(\Vec{\eta}_j), 
 \  (i= 2 \sim (N-1)).
\end{eqnarray}
This relation is equivalent to the relation
\begin{eqnarray}
 \Vec{\xi}_i = \sum_{j=2}^{N-1} d_{ij} \Vec{\eta}_j, \  (i= 2 \sim (N-1)). 
 \label{eq:xi_d_eta}
\end{eqnarray}
By inserting Eq.~(\ref{eq:xi_d_eta}) into Eq.~(\ref{eq:Jacobi_quadra}) we obtain 
\begin{eqnarray}
 \sum_{i=1}^N \Vec{r}_i^2 &=& N \Vec{R}^2 + \sum_{i=1}^{N-1} 
    \frac{\nu_i}{\nu} \Vec{\xi}_i^2   \\ 
  &=& N \Vec{R}^2 + \frac{\nu_1}{\nu} \Vec{\eta}_1^2 + 
   \sum_{i,i'= 2}^{N-1} a_{i,i'} \Vec{\eta}_i \cdot \Vec{\eta}_{i'}, 
   \label{eq:ortho_quadra} \\ 
 a_{i,i'} &=& \sum_{j=2}^{N-1} \frac{\nu_j}{\nu} d_{j,i} d_{j,i'}.  
\end{eqnarray}
It is to be noticed that unlike the Jacobi coordinates the mutual 
orthogonality within the $(N-2)$ coordinates 
$\{ \Vec{\eta}_i, \ i= 2 \sim (N-1) \}$ does not hold in general. 

By using Eq.~(\ref{eq:ortho_quadra}) we can calculate the internal 
one-particle density matrix $\rho_{\rm int}^{(1)}$ as follows 
\begin{eqnarray}
 \rho_{\rm int}^{(1)}(\Vec{\eta},\Vec{\eta}^\prime) &\propto& 
   \int d\Vec{\eta}_2 \cdots d\Vec{\eta}_{N-1} 
   \exp\left[ -\nu \left( \frac{\nu_1}{\nu}  \Vec{\eta}^2 + \sum_{i,i'= 2}^{N-1} a_{i,i'} \Vec{\eta}_i \cdot \Vec{\eta}_{i'} \right)\right] 
   \nonumber \\ 
 && \times \exp\left[ -\nu \left( \frac{\nu_1}{\nu} {\Vec{\eta}^\prime}^2 + \sum_{i,i'= 2}^{N-1} a_{i,i'} \Vec{\eta}_i \cdot \Vec{\eta}_{i'} \right) \right] \\ 
 &\propto& \exp\left[ - \frac{N-1}{N} \nu \left( \Vec{\eta}^2 + {\Vec{\eta}^\prime}^2 \right) \right]. 
\end{eqnarray}
This result shows that $\rho_{\rm int}^{(1)}(\Vec{\eta},\Vec{\eta}^\prime)$ 
 is just the same as the Jacobi-type internal one-particle density matrix. 
Thus we see that our above requirement concerning the internal coordinates, 
 gives us a {\it unique result} for the internal one-particle density matrix. 
It is just identical to the Jacobi-type internal one-particle density matrix. 

In the previous section \ref{subsec:convergence_BEC_macro}
 we have proposed also another requirement, namely 
 that the internal one-particle density matrix should converge to the one-particle density 
 matrix in the laboratory frame in the macroscopic limit.  
As we discussed in Sec.~\ref{subsec:convergence_BEC_macro}, this requirement
 implies that there should not 
 appear any cross terms of $\Vec{\xi}_1$ with the other internal 
 coordinates in the internal wave function at least in the macroscopic limit.   
Clearly the absence of the cross terms of 
 $\Vec{\xi}_1$ with the other internal coordinates is realized only 
 when the coordinate $\Vec{\xi}_1$ is orthogonal to all the other 
 $(N-2)$ internal coordinates. 
Thus the criterion in the 
 preceding subsection results from the present requirement of the 
 orthogonality of $\Vec{\xi}_1$ to the other $(N-2)$ internal coordinates, where the $( N - 2 )$ coordinates
 generally do not need to be mutually orthogonal. 

One may argue that a physical quantity should not depend on the choice of 
 the coordinate system. 
In the case of the one-particle density matrix this 
 argument can be true under the condition that we have extracted maximum 
 information on the one-particle degree of freedom of the system.  
As already noticed this condition is the same as the requirement that the 
 coordinate of the one-particle density matrix is orthogonal to all the other 
 coordinates of the system. 
This is also true in the macroscopic system where 
 we usually adopt the coordinate system composed of individual particle 
 coordinates $\{ \Vec{r}_i, \ (i= 1 \sim N) \}$.  
The coordinate $\Vec{r}_1$ 
 of the one-particle density matrix is of course orthogonal to all the other 
 coordinates. 
If we adopt the coordinate system $\{ \Vec{R}, \Vec{q}_i, \ (i= 1 \sim (N-1)) \}$
 and calculate the one-particle density matrix with 
 respect to the coordinate $\Vec{q}_1$ which is practically the same as 
 $\Vec{r}_1$ in the macroscopic system by integrating out the remaining 
 coordinates $\{ \Vec{R}, \Vec{q}_i, \ (i= 2 \sim (N-1)) \}$, we get the 
 result given in Eq.~(\ref{eq:one_body_den_int_infinite}), namely 
 $(2\nu/\pi)^{3/2} \exp [-(3/2) \nu(\Vec{q}_1^2+{\Vec{q}_1^\prime}^2) + 
 \nu \Vec{q}_1 \cdot \Vec{q}_1^\prime ]$.  
This inadequate result is, of course, due to the non-orthogonality of $\Vec{q}_1$ to the other coordinates 
 $\{ \Vec{q}_i, \ (i= 2 \sim (N-1)) \}$, as we have already seen earlier. 

The PP-type internal one-particle density matrix $\rho_{\rm int,PP}^{(1)}$ 
 is different from the Jacobi-type one $\rho_{\rm int,J}^{(1)}$. 
This is true 
 for the non-diagonal elements but for the diagonal elements they are the 
 same except for the Jacobian factor 
 $\partial(\Vec{q}_1)/\partial(\Vec{\xi}_1) = ((N-1)/N)^3$,
 (which goes to unity in the macroscopic limit)
\begin{eqnarray}
 \rho_{\rm int,PP}^{(1)}(\Vec{q}_1,\Vec{q}_1) = \left(\frac{N}{N-1}\right)^3 
  \rho_{\rm int,J}^{(1)}(\Vec{\xi}_1,\Vec{\xi}_1). 
  \label{eq:PPJ}
\end{eqnarray} 
It is to be noticed that both $\rho_{\rm int,PP}^{(1)}$ and 
$\rho_{\rm int,J}^{(1)}$ are normalized, 
\begin{eqnarray}
 \int d\Vec{q}_1 \rho_{\rm int,PP}^{(1)}(\Vec{q}_1,\Vec{q}_1) = 
 \int d\Vec{\xi}_1 \rho_{\rm int,J}^{(1)} (\Vec{\xi}_1,\Vec{\xi}_1) = 1. 
\end{eqnarray} 
The equality of Eq.~(\ref{eq:PPJ}) is easily proved by using 
 Eqs.~(\ref{eq:one_body_den_int}) and (\ref{eq:Jacobi_one_body_den_int}). 
Let us study these relations between $\rho_{\rm int,PP}^{(1)}$ and 
 $\rho_{\rm int,J}^{(1)}$ a little more in detail.  First we note the 
 relation between the two coordinate systems, 
 $\{ \Vec{q}_i, \ (i= 1 \sim (N-1)) \}$ and 
 $\{ \Vec{\xi}_i, \ (i= 1 \sim (N-1)) \}$, 
\begin{eqnarray}
 \Vec{q}_i = - \sum_{j=1}^{i-1} \frac{1}{N+1-j} \Vec{\xi}_j + 
   \frac{N-1}{N+1-i} \Vec{\xi}_i. 
\end{eqnarray} 
When we calculate $\rho_{\rm int,J}^{(1)}$, we make the following 
integration 
\begin{eqnarray}
 \rho_{\rm int,J}^{(1)}(\Vec{\xi}_1,\Vec{\xi}_1^\prime) &=& 
   \int d\Vec{\xi}_2 \cdots d\Vec{\xi}_{N-1}~\Phi_{\rm int}(\Vec{\xi}_1,
    \Vec{\xi}_2,\cdots,\Vec{\xi}_{N-1}) \Phi_{\rm int}(\Vec{\xi}_1^\prime,
    \Vec{\xi}_2,\cdots,\Vec{\xi}_{N-1}) \\
  &=& \frac{\partial(\Vec{\xi}_2,\cdots,\Vec{\xi}_{N-1})}
      {\partial(\Vec{q}_2,\cdots,\Vec{q}_{N-1})} 
   \int d\Vec{q}_2 \cdots d\Vec{q}_{N-1}~{\widehat \Phi}_{\rm int}(\Vec{q}_1,
    \Vec{q}_2,\cdots,\Vec{q}_{N-1}) \nonumber \\ 
  && \hspace*{2cm} \times {\widehat \Phi}_{\rm int}(\Vec{q}_1^\prime,
    \Vec{q}_2^\prime,\cdots,\Vec{q}_{N-1}^\prime) \\
 \Vec{q}_i^\prime &=& \Vec{q}_i(\Vec{\xi}_1 \rightarrow \Vec{\xi}_1^\prime) 
    =  \Vec{q}_i + \frac{1}{N} (\Vec{\xi}_1 - \Vec{\xi}_1^\prime), 
\end{eqnarray} 
where 
\begin{eqnarray}
  &&{\widehat \Phi}_{\rm int}(\Vec{q}_1,\Vec{q}_2,\cdots,\Vec{q}_{N-1}) = 
    \Phi_{\rm int}(\Vec{\xi}_1,\Vec{\xi}_2,\cdots,\Vec{\xi}_{N-1}), \\ 
  &&{\widehat \Phi}_{\rm int}(\Vec{q}_1^\prime,\Vec{q}_2^\prime,\cdots,
    \Vec{q}_{N-1}^\prime) = \Phi_{\rm int}(\Vec{\xi}_1^\prime,
    \Vec{\xi}_2,\cdots,\Vec{\xi}_{N-1}). 
\end{eqnarray}
For non-diagonal elements of $\rho_{\rm int,J}^{(1)}(\Vec{\xi}_1,
\Vec{\xi}_1^\prime)$, since $\Vec{\xi}_1 \ne \Vec{\xi}_1^\prime$, it 
follows that $\Vec{q}_i^\prime \ne \Vec{q}_i$ for all $i = 1 \sim (N-1)$. 
This tells us that $\rho_{\rm int,J}^{(1)}(\Vec{\xi}_1,\Vec{\xi}_1^\prime)$ 
can not be proportional to $\rho_{\rm int,PP}^{(1)}(\Vec{q}_1,
\Vec{q}_1^\prime)$ because the latter is obtained from 
\begin{eqnarray}
  \int d\Vec{q}_2 \cdots d\Vec{q}_{N-1}~{\widehat \Phi}_{\rm int}
    (\Vec{q}_1,\Vec{q}_2,\cdots,\Vec{q}_{N-1}) \ 
    {\widehat \Phi}_{\rm int}(\Vec{q}_1^\prime, \Vec{q}_2,\cdots,
    \Vec{q}_{N-1}). 
\end{eqnarray}
On the other hand, for diagonal elements, $\rho_{\rm int,J}^{(1)}
(\Vec{\xi}_1,\Vec{\xi}_1)$, since $\Vec{\xi}_1 = \Vec{\xi}_1^\prime$, there 
follows $\Vec{q}_i^\prime = \Vec{q}_i$ for all $i = 1 \sim (N-1)$. 
This tells us that $\rho_{\rm int,J}^{(1)}(\Vec{\xi}_1,\Vec{\xi}_1)$ 
is now proportional to $\rho_{\rm int,PP}^{(1)}(\Vec{q}_1,\Vec{q}_1)$. 
The appearance of the term $(1/N) (\Vec{\xi}_1 - \Vec{\xi}_1^\prime)$ 
in the relation $\Vec{q}_i^\prime = \Vec{q}_i + (1/N) (\Vec{\xi}_1 - 
\Vec{\xi}_1^\prime)$ just stems from the non-orthogonality of 
$\Vec{q}_1$ to $\Vec{q}_i \ (i=2 \sim (N-1))$. 

We have proved in this section that the internal one-body density matrix is uniquely determined
 for the $0S$ harmonic oscillator wave function.
However, it is noted that this uniqueness holds in the case of a general wave function including the $0S$ harmonic oscillator one. 
In fact, Suzuki et al.~\cite{Suzuki1,Suzuki2} have already given the proof of the uniqueness for the more general wave
 function $\Psi$ which is expanded in terms of the correlated Gaussian basis $g$. 
The explicit forms of $\Psi$ and $g$ are expressed as follows:
\begin{eqnarray}
&&\Psi=\sum_k C_k {\cal S} g(\Vec{s}^{(k)};A^{(k)},\Vec{x}) \label{eq:SVM}\\ 
&&g=g(\Vec{s};A,\Vec{x})=\exp\Big[ -\frac{1}{2}\sum_{i,j}A_{ij}\Vec{x}_i \cdot \Vec{x}_j + \sum_i \Vec{s}_i \cdot \Vec{x}_i \Big], 
\end{eqnarray} 
where $\Vec{x}=\{\Vec{x}_1,\Vec{x}_2,\cdots,\Vec{x}_{N-1}\}$
 and ${\cal S}$ are a set of internal coordinate of $N$-boson system and the symmetrization
 operator acting on the $N$ bosons, respectively. 
$C_k$, $A^{(k)}$ and $\Vec{s}^{(k)}$ are the expansion parameters,
 in which $A^{(k)}$ denotes a symmetric positive-defined $(N-1)\times(N-1)$ matrix,
 and $\Vec{s}^{(k)}$ represents $\Vec{s}^{(k)}=\{\Vec{s}_1^{(k)},\Vec{s}_2^{(k)},\cdots,\Vec{s}_{N-1}^{(k)}\}$.
The correlated Gaussian basis $g$ is often used in ab-initio calculations and has succeeded
 in describing structures of many few-body systems~\cite{Suzuki3}. 
The uniqueness of the internal one-body density matrix $\rho$ for the wave function $\Psi$ in Eq.~(\ref{eq:SVM}) is 
 proved~\cite{Suzuki1,Suzuki2} under the condition that one takes the following
 set of coordinates $\Vec{y}=\{\Vec{y}_1,\Vec{y}_2,\cdots,\Vec{y}_{N-1}\}$ (obtained by a linear transformation from the coordinates $\Vec{x}$)
 as the coordinates of $\Psi$ adopted in calculating $\rho$:~$\Vec{q}_1$
 is chosen as the coordinate ($\Vec{y_1}$) of the internal one-body density matrix and the remaining 
 $(N-2)$ internal coordinates ($\Vec{y}_2,\Vec{y}_3,\cdots,\Vec{y}_{N-1}$)
 are orthogonal to $\Vec{q}_1$, although the $(N-2)$ coordinates do not need to be mutually orthogonal. 
This requirement for the coordinates $\Vec{y}$ just corresponds to the "maximally independence
 of the coordinate $\Vec{\xi}_1$ ($\Vec{q}_1$) from the rest of $(N-2)$ internal coordinates" as mentioned above. 
The proof we gave in this section is specified to the simple system which is composed of $N$ bosons
 in the harmonic oscillator potential and has been discussed by many authors
 (for example, see Refs.~\cite{PP,wilkin,arxiv,letter}).

\section{Summary}

We investigated  the internal one-particle density matrix 
in the case of ideal Bose-Einstein condensates with a finite number 
( $N$ ) of particles in a harmonic trap.  We calculated 
the explicit form of the internal one-particle density matrix 
following the definition of Pethick and Pitaevskii  (PP) and 
solved  its eigenvalue problem. The result was found to show 
a fragmented condensate, contrary to what PP 
expected. On the other hand the Jacobi-type  internal one-particle 
density matrix gives us complete condensation.  
It means that the internal one-particle density matrix is
 different in general for different choices of the internal coordinate system. 
In this paper we outlined two physically motivated criteria for
 the choice of the adequate 
 coordinate system leading to a unique answer for the internal 
 one-particle density matrix. 
One criterion is that in the infinite 
 particle number ( $N=\infty$ ) limit the internal one-particle density 
 matrix should have the same eigenvalues and eigenfunctions as 
 those of the corresponding ideal Bose-Einstein condensate in the 
 laboratory frame. 
The other criterion is that the coordinate of the 
 internal one-particle density matrix which is either 
 $\Vec{q}_1 = \Vec{r}_1 - \Vec{R}$ or $\Vec{\xi}_1 = (N/(N-1)) \Vec{q}_1$,  
 should be maximally independent from  the remaining $( N - 2 )$ 
 internal coordinates.  
Mathematically this criterion means that $\Vec{q}_1$ (or $\Vec{\xi}_1$)
 is orthogonal to the remaining  $( N - 2 )$ internal coordinates,
 though the $( N - 2 )$ coordinates, in general, do not need to be mutually orthogonal.
This second criterion was shown to imply the first criterion. 
We saw that the internal Jacobi 
 coordinate system satisfies these two criteria while the internal 
 coordinate system adopted by Pethick and Pitaevskii for the 
 construction of the internal one-particle density matrix does not. 
Furthermore we argued that these two criteria {\it uniquely} determine  
 the internal one-particle density matrix which is the same as that 
 calculated with the Jacobi coordinates. 
The results of this paper justify the use of the Jacobi 
 coordinates in Refs.~\cite{matsumura,yamada} where the Bose 
 condensation of a few $\alpha$-particles in extended states of self-conjugate 
 light nuclei was considered. 
However, our results are of more general 
 interest. For example the number of bosons captured in each site of an optical 
 lattice is often very small~\cite{zwerg} and, therefore, our 
 analysis surely applies to that situation as well. We also believe that our 
 present study has considerably clarified the somewhat controversial issue of 
 how to define a Bose condensate of a finite number of particles in their 
 internal coordinate system. 
This question is particularly, but not only, 
 relevant for selfbound Bose systems, like a loosely bound gas of $\alpha$-particles
 in nuclei or nano-droplets of liquid $^{4}$He.



The authors thank Kanto Gakuin University (KGU) and Yukawa Institute for Theoretical
 Physics at Kyoto University, Japan. 
Discussions during the KGU Yokohama Autumn School of Nuclear Physics and the YIPQS
 international molecule workshop held in October 2008
 were useful to complete this work. 

\appendix

\section{Solution of the Eigenvalue Problem of the Symmetric Gaussian 
Integral Kernel}\label{app:A}

The one-particle density matrices discussed in this paper are symmetric 
Gaussian integral kernels of the form 
\begin{eqnarray}
 \langle \Vec{r} | \hat{Q} | \Vec{r}^\prime \rangle = 
 Q(\Vec{r}, \Vec{r}^\prime) = \left( \frac{2a-b}{\pi} \right)^{\frac{3}{2}} 
   \exp \{ -a(\Vec{r}^2 + {\Vec{r}^\prime}^2) + b \Vec{r} \cdot 
   \Vec{r}^\prime \}. 
 \label{eq:qdef}
\end{eqnarray}
This kernel is normalized 
\begin{eqnarray}
 {\rm Tr}~\hat{Q} = \int d\Vec{r} Q(\Vec{r}, \Vec{r}) = 1,
\end{eqnarray}
for which we need the condition $2a-b > 0$. 

The eigenvalue problem of the kernel $Q(\Vec{r}, \Vec{r}^\prime)$ can be 
 solved analytically. 
There are several ways to solve it. 
We can e.g.~use the method given in Ref.~\cite{gajda}. 
Here we explain another procedure which gives us the following answer (see below)
\begin{eqnarray}
  && \hat{Q} = \left( \frac{2a-b}{a+c} \right)^{\frac{3}{2}} \left( \frac{2(a-c)}
   {b} \right)^{\Vec{\alpha}_c^\dagger \cdot \Vec{\alpha}_c}, \ \ 
   c = \sqrt{a^2-\frac{b^2}{4}}, 
   \label{eq:qegv}   \\ 
  && \Vec{\alpha}_c^\dagger = \sqrt{c} \left( \Vec{r} - \frac{1}{2c} 
   \frac{\partial}{\partial \Vec{r}} \right), \ \ 
   \Vec{\alpha}_c = \sqrt{c} \left( \Vec{r} + \frac{1}{2c} 
   \frac{\partial}{\partial \Vec{r}} \right).     
\end{eqnarray}
Here $\Vec{\alpha}_c^\dagger$ and $\Vec{\alpha}_c$ are creation and 
annihilation operators of harmonic oscillation of size parameter $c$, 
respectively. The eigenfunctions of $\hat{Q}$ are just the same as those of the 
operator $\Vec{\alpha}_c^\dagger \cdot \Vec{\alpha}_c$ which are of course 
the harmonic oscillator functions of the size parameter $c$. Thus we have 
\begin{eqnarray}
 && \hat{Q} | \phi_{nLM}(c) \rangle = \lambda_n^{(L,M)} | \phi_{nLM}(c) \rangle, \\ 
 && \langle \Vec{r}|\phi_{nLM}(c) \rangle = \phi_{nLM}(\Vec{r}, c), \\ 
 && \lambda_n^{(L,M)} = \left( \frac{2a-b}{a+c} \right)^{\frac{3}{2}} 
   \left( \frac{2(a-c)}{b} \right)^{2n+L}.
\end{eqnarray}
The explicit form of $\phi_{nLM}(\Vec{r}, c)$ is given in Eq.~(\ref{eq:harmosci}). 
It is to be noted that the eigenvalue $\lambda_n^{(L,M)}$ depends only on 
the number of harmonic oscillator quanta $(2n+L)$. It is just the SU(3)-scalar 
property of the operator $\hat{Q}$. 
A merit of our method as explained below, is that it gives us a clearer
 understanding of the SU(3) symmetry of the operator $\hat{Q}$. 

The relation of Eq.~(\ref{eq:qegv}) comes directly from the following 
theorem~\cite{hawy} valid for a general operator $P$: 
\begin{eqnarray}
 P = {\cal N}_{\rm op} \{ p(\Vec{\alpha}_\gamma^\dagger, 
 \Vec{\alpha}_\gamma) \}, 
 \label{eq:general}
\end{eqnarray}
where ${\cal N}_{\rm op}$ is the operator of normal ordering, and 
\begin{eqnarray}
 \Vec{\alpha}_\gamma^\dagger &=& \sqrt{\gamma} \left( \Vec{r} - 
    \frac{1}{2\gamma} \frac{\partial}{\partial \Vec{r}} \right), \ \ 
   \Vec{\alpha}_\gamma \ = \ \sqrt{\gamma} \left( \Vec{r} + 
    \frac{1}{2\gamma} \frac{\partial}{\partial \Vec{r}} \right), \\ 
 p(\Vec{Z}^*, \Vec{Z}) &=& \frac{ \langle A_\gamma(\Vec{Z}) | P | 
   A_\gamma(\Vec{Z}^\prime) \rangle }{ \langle A_\gamma(\Vec{Z}) | 
   A_\gamma(\Vec{Z}^\prime) \rangle } \\ 
 &=& \exp( -\Vec{Z}^* \cdot \Vec{Z}^\prime ) \langle A_\gamma(\Vec{Z}) 
   | P | A_\gamma(\Vec{Z}^\prime) \rangle,  \\ 
 \langle \Vec{r}| A_\gamma(\Vec{Z}) \rangle &=&  A_\gamma(\Vec{r},\Vec{Z}) 
   \ = \ \left( \frac{2\gamma}{\pi} \right)^{\frac{3}{4}} \exp \left[ 
   -\gamma \left( \Vec{r} - \frac{\Vec{Z}}{\sqrt{\gamma}} \right)^2 + 
   \frac{1}{2} \Vec{Z}^2 \right].    
\end{eqnarray}
The state $|A_\gamma(\Vec{Z})\rangle$ is the well-known coherent state 
of harmonic oscillator of size parameter $\gamma$, 
\begin{eqnarray}
  |A_\gamma(\Vec{Z})\rangle &=& \exp \{ \Vec{Z} \cdot 
     \Vec{\alpha}_\gamma^\dagger \}\ |A_\gamma(\Vec{Z}=0)\rangle,  \\ 
  \Vec{\alpha}_\gamma |A_\gamma(\Vec{Z})\rangle &=& \Vec{Z} 
    |A_\gamma(\Vec{Z})\rangle, 
    \label{eq:coherent}   \\ 
  A_\gamma(\Vec{r},\Vec{Z}) &=& A_\gamma(x,Z_x) A_\gamma(y,Z_y) 
    A_\gamma(z,Z_z), \\ 
  A_\gamma(x,Z_x) &=& \left( \frac{2\gamma}{\pi} \right)^{\frac{1}{4}} 
    \exp \left[ -\gamma \left( x - \frac{Z_x}{\sqrt{\gamma}} \right)^2 + 
   \frac{1}{2} Z_x^2 \right]   \\ 
  &=& \sum_{n=0}^\infty \frac{(Z_x)^n}{\sqrt{n!}} X_n(x,\gamma), 
  \label{eq:expand} \\ 
  X_n(x,\gamma) &=& \frac{(\alpha^\dagger_{\gamma,x})^n}{\sqrt{n!}} 
    A_\gamma(x,Z_x=0). 
\end{eqnarray}
The proof of the general theorem of Eq.~(\ref{eq:general}) is quite easy.  
It is given by comparing the matrix elements of two operators, $P$ and 
 ${\cal N}_{\rm op} \{ p(\Vec{\alpha}_\gamma^\dagger, \Vec{\alpha}_\gamma) \}$, 
 formed with the coherent states $|A_\gamma(\Vec{Z})\rangle$ which constitute 
 an overcomplete set of states. 
This is shown by the following operation:
\begin{eqnarray} 
 \langle A_\gamma(\Vec{Z}) | {\cal N}_{\rm op} \{ 
   p(\Vec{\alpha}_\gamma^\dagger, 
   \Vec{\alpha}_\gamma) \} | A_\gamma(\Vec{Z}^\prime) \rangle &=& 
   p(\Vec{Z}^*, \Vec{Z}^\prime) \langle A_\gamma(\Vec{Z}) | 
   A_\gamma(\Vec{Z}^\prime) \rangle 
   \label{eq:matelem}   \\ 
  &=& \langle A_\gamma(\Vec{Z}) | P |A_\gamma(\Vec{Z}^\prime) \rangle.  
\end{eqnarray}
The equality of Eq.~(\ref{eq:matelem}) is due to Eq.~(\ref{eq:coherent}). 

By applying the above general theorem to our present operator $Q$ of 
Eq.~(\ref{eq:qdef}), we obtain  
\begin{eqnarray}
 q(\Vec{Z}^*, \Vec{Z}) &=& \frac{ \langle A_\gamma(\Vec{Z}) | Q | 
    A_\gamma(\Vec{Z}^\prime) \rangle }{ \langle A_\gamma(\Vec{Z}) | 
    A_\gamma(\Vec{Z}^\prime) \rangle } \\ 
  &=& \left( \frac{\pi^2}{(\gamma + a)^2 - (b^2/4)} \right)^{\frac{3}{2}} 
    \exp \left[ F({\Vec{Z}^*}^2 + {\Vec{Z}^\prime}^2) + G \Vec{Z}^* \cdot 
    \Vec{Z}^\prime \right], \\ 
  F &=& -\frac{1}{2} + \frac{\gamma(\gamma+a)}{(\gamma + a)^2 - (b^2/4)}, \\ 
  G &=& -1 + \frac{\gamma b}{(\gamma + a)^2 - (b^2/4)}.   
\end{eqnarray}
By choosing 
\begin{eqnarray}
 \gamma = c = \sqrt{a^2-\frac{b^2}{4}}, 
\end{eqnarray}
we obtain $F=0$.  Thus we have for $\gamma = c$ 
\begin{eqnarray}
 q(\Vec{Z}^*, \Vec{Z}^\prime) = \left( \frac{2a-b}{a+c} \right)^{\frac{3}{2}} 
   \exp \left[ \left( \frac{2(a-c)}{b} - 1 \right) \Vec{Z}^* \cdot 
   \Vec{Z}^\prime \right]. 
\end{eqnarray}
By using the well-known formula~\cite{wilcox} 
\begin{eqnarray}
  {\cal N}_{\rm op} \{ \exp [ (Y-1) \Vec{\alpha}_\gamma^\dagger \cdot 
   \Vec{\alpha}_\gamma ] \} = Y^{ \Vec{\alpha}_\gamma^\dagger \cdot 
   \Vec{\alpha}_\gamma },
   \label{eq:wellknown} 
\end{eqnarray}
we obtain the desired result of Eq.~(\ref{eq:qegv}) 
\begin{eqnarray}
 Q &=& {\cal N}_{\rm op} \{ q( \Vec{\alpha}_c^\dagger,  \Vec{\alpha}_c) \} \\ 
   &=& \left( \frac{2a-b}{a+c} \right)^{\frac{3}{2}} \left( \frac{2(a-c)}
   {b} \right)^{\Vec{\alpha}_c^\dagger \cdot \Vec{\alpha}_c}.
\end{eqnarray}
The formula of Eq.~(\ref{eq:wellknown}) can also easily be proved by comparing 
the matrix elements of $ {\cal N}_{\rm op} \{ \exp [ (Y-1) 
\Vec{\alpha}_\gamma^\dagger \cdot \Vec{\alpha}_\gamma ] \}$ and 
$Y^{ \Vec{\alpha}_\gamma^\dagger \cdot \Vec{\alpha}_\gamma }$ formed with 
coherent 
states. For treating $Y^{ \Vec{\alpha}_\gamma^\dagger \cdot 
\Vec{\alpha}_\gamma }$, the expansion of the coherent state by harmonic 
oscillator functions given in Eq.~(\ref{eq:expand}) is useful.

\section{Bosonic Symmetry of the Jacobi-Type Internal One-Particle Density Matrix}\label{app:B}

The bosonic symmetry of the Jacobi-type internal one-particle density 
matrix defined in Eq.~(\ref{eq:Jacobi_one_body_den_int_general}) can be 
seen in the following identity relation, 
\begin{eqnarray}
 \rho_{\rm int,J}^{(1)}(\Vec{\xi},\Vec{\xi}^\prime) &=& 
  \int d\Vec{\xi}_2 \cdots d\Vec{\xi}_{N-1}~\Phi_{\rm int}(\Vec{\xi},
   \Vec{\xi}_2,\cdots,\Vec{\xi}_{N-1}) \Phi_{\rm int}(\Vec{\xi}^\prime,
   \Vec{\xi}_2,\cdots,\Vec{\xi}_{N-1}) \\ 
 &=& \int d\Vec{\xi}_1 d\Vec{\xi}_2 \cdots d\Vec{\xi}_{N-1} \times 
   d\Vec{\xi}_1^\prime d\Vec{\xi}_2^\prime \cdots d\Vec{\xi}_{N-1}^\prime \nonumber \\ 
 && \quad \times \Phi_{\rm int}(\Vec{\xi}_1,\Vec{\xi}_2,\cdots,\Vec{\xi}_{N-1}) 
  {\widehat O}(1,\Vec{\xi},\Vec{\xi}^\prime) \Phi_{\rm int}
  (\Vec{\xi}_1^\prime,\Vec{\xi}_2^\prime,\cdots,\Vec{\xi}_{N-1}^\prime) \\  
 &=& \int d\Vec{\xi}_1 d\Vec{\xi}_2 \cdots d\Vec{\xi}_{N-1}  \times 
  d\Vec{\xi}_1^\prime d\Vec{\xi}_2^\prime \cdots d\Vec{\xi}_{N-1}^\prime \nonumber \\ 
 && \quad \times \Phi_{\rm int}(\Vec{\xi}_1,\Vec{\xi}_2,\cdots,\Vec{\xi}_{N-1}) 
  \frac{1}{N} \sum_{k=1}^N {\widehat O}(k,\Vec{\xi},\Vec{\xi}^\prime) 
  \Phi_{\rm int}(\Vec{\xi}_1^\prime,\Vec{\xi}_2^\prime,\cdots,
  \Vec{\xi}_{N-1}^\prime),  \label{eq:symmetry}  \\ 
 {\widehat O}(k,\Vec{\xi},\Vec{\xi}^\prime) &=& \delta(\Vec{\xi}_1^k - 
    \Vec{\xi}) \delta({\Vec{\xi}_1^k}^\prime - \Vec{\xi}^\prime) 
    \prod_{j \ge 2} \delta(\Vec{\xi}_j^k - {\Vec{\xi}_j^k}^\prime). 
\end{eqnarray}
Here we define $N$ different sets of Jacobi coordinates 
$(\Vec{\xi}_1^k, \Vec{\xi}_2^k, \cdots, \Vec{\xi}_{N-1}^k)$ 
by $N$ cyclic permutations of particle indices $(1, 2, \cdots, N)$ 
\begin{eqnarray}
 && \Vec{\xi}_i^k = \Vec{r}_{p_k(i)} - \frac{1}{N-i} \sum_{j=i+1}^N 
    \Vec{r}_{p_k(j)}, \\ 
 && (p_k(1), p_k(2),\cdots, p_k(N)) = (k, k+1,\cdots, N, 1, 2,\cdots, k-1), \ \ \quad (k = 1 \sim N ). 
\end{eqnarray}
We should note that, since 
$\Phi_{\rm int}(\Vec{\xi}_1,\Vec{\xi}_2,\cdots,\Vec{\xi}_{N-1})$ is 
totally symmetric for particle permutations as is clear from the relation 
$\Phi(\Vec{r}_1,\Vec{r}_2,\cdots,\Vec{r}_N) = \Phi_{\rm int}(\Vec{\xi}_1,
\Vec{\xi}_2,\cdots,\Vec{\xi}_{N-1}) \Phi_{\rm cm}(\Vec{R})$, we have 
\begin{eqnarray}
  \Phi_{\rm int}(\Vec{\xi}_1,\Vec{\xi}_2,\cdots,\Vec{\xi}_{N-1}) = 
  \Phi_{\rm int}(\Vec{\xi}_1^k,\Vec{\xi}_2^k,\cdots,\Vec{\xi}_{N-1}^k) 
 \quad \quad (k = 1 \sim N ).  \label{eq:symm1}
\end{eqnarray}
And of course we have 
\begin{eqnarray}
  d\Vec{\xi}_1 d\Vec{\xi}_2 \cdots d\Vec{\xi}_{N-1} = 
  d\Vec{\xi}_1^k d\Vec{\xi}_2^k \cdots d\Vec{\xi}_{N-1}^k. \label{eq:symm2} 
\end{eqnarray}
The expression of Eq.~(\ref{eq:symmetry}) is the manifestation of the 
 bosonic symmetry of the Jacobi-type internal one-particle density matrix. 
This equality is due to the following relations 
\begin{eqnarray}
 \langle {\widehat O}(k,\Vec{\xi},\Vec{\xi}^\prime) \rangle = 
 \langle {\widehat O}(1,\Vec{\xi},\Vec{\xi}^\prime) \rangle, \quad 
 (k = 2 \sim N ), \label{eq:symmetrya}, 
\end{eqnarray}  
where $\langle {\widehat O}(k,\Vec{\xi},\Vec{\xi}^\prime) \rangle$ is 
defined as 
\begin{eqnarray}
 \langle {\widehat O}(k,\Vec{\xi},\Vec{\xi}^\prime) \rangle &=&  
 \int d\Vec{\xi}_1 d\Vec{\xi}_2 \cdots d\Vec{\xi}_{N-1}  \times  
  d\Vec{\xi}_1^\prime d\Vec{\xi}_2^\prime \cdots d\Vec{\xi}_{N-1}^\prime \nonumber \\ 
 && \quad \times \Phi_{\rm int}(\Vec{\xi}_1,\Vec{\xi}_2,\cdots,\Vec{\xi}_{N-1}) 
  {\widehat O}(k,\Vec{\xi},\Vec{\xi}^\prime) 
  \Phi_{\rm int}(\Vec{\xi}_1^\prime,\Vec{\xi}_2^\prime,\cdots,
  \Vec{\xi}_{N-1}^\prime),  \\ 
 && \quad (k = 1 \sim N ). \nonumber
\end{eqnarray}

We can prove the equality of Eq.~(\ref{eq:symmetrya}), by using  
 Eq.~(\ref{eq:symm1}) and Eq.~(\ref{eq:symm2}), as follows  
\begin{eqnarray}
 \langle {\widehat O}(k,\Vec{\xi},\Vec{\xi}^\prime) \rangle &=&  
 \int d\Vec{\xi}_1 d\Vec{\xi}_2 \cdots d\Vec{\xi}_{N-1}  \times  
  d\Vec{\xi}_1^\prime d\Vec{\xi}_2^\prime \cdots d\Vec{\xi}_{N-1}^\prime \nonumber \\ 
 && \quad \times \Phi_{\rm int}(\Vec{\xi}_1,\Vec{\xi}_2,\cdots,\Vec{\xi}_{N-1}) 
  {\widehat O}(k,\Vec{\xi},\Vec{\xi}^\prime) 
  \Phi_{\rm int}(\Vec{\xi}_1^\prime,\Vec{\xi}_2^\prime,\cdots,
  \Vec{\xi}_{N-1}^\prime) \\ 
 &=& \int d\Vec{\xi}_1^k d\Vec{\xi}_2^k \cdots d\Vec{\xi}_{N-1}^k  \times  
  d{\Vec{\xi}_1^k}^\prime d{\Vec{\xi}_2^k}^\prime \cdots 
  d{\Vec{\xi}_{N-1}^k}^\prime \nonumber \\ 
 && \quad \times \Phi_{\rm int}(\Vec{\xi}_1^k,\Vec{\xi}_2^k,\cdots,\Vec{\xi}_{N-1}^k) 
  {\widehat O}(k,\Vec{\xi},\Vec{\xi}^\prime) 
  \Phi_{\rm int}({\Vec{\xi}_1^k}^\prime,{\Vec{\xi}_2^k}^\prime,\cdots,
  {\Vec{\xi}_{N-1}^k}^\prime) \\ 
 &=& \int d\Vec{\xi}_1 d\Vec{\xi}_2 \cdots d\Vec{\xi}_{N-1}  \times  
  d\Vec{\xi}_1^\prime d\Vec{\xi}_2^\prime \cdots d\Vec{\xi}_{N-1}^\prime \nonumber \\ 
 && \quad \times \Phi_{\rm int}(\Vec{\xi}_1,\Vec{\xi}_2,\cdots,\Vec{\xi}_{N-1}) 
  {\widehat O}(1,\Vec{\xi},\Vec{\xi}^\prime) 
  \Phi_{\rm int}(\Vec{\xi}_1^\prime,\Vec{\xi}_2^\prime,\cdots,
  \Vec{\xi}_{N-1}^\prime) \\    
 &=& \langle {\widehat O}(1,\Vec{\xi},\Vec{\xi}^\prime) \rangle. 
\end{eqnarray}


\end{document}